\documentclass[12pt]{article}    
\usepackage{aaspp4}              
\usepackage{flushrt}
\usepackage{graphicx}
\DeclareGraphicsExtensions{.ps,.jpg,.pdf,.mps,.png}

\begin{document}
\slugcomment{Accepted for Publication in The Astrophysical Journal}
\lefthead{Courteau {\it et al.}}
\righthead{Shellflow Survey}

\newcommand{\be}{\begin{equation}}
\newcommand{\ee}{\end{equation}}

\def\degs{\ifmmode^\circ\else$^\circ$\fi}
\def\heion{\ion{He}{2}}
\hyphenation{an-is-o-tro-pies}
\hyphenation{sam-ples}
\hyphenation{an-is-o-tro-py}
\def \kms{\ifmmode\,{\rm km}\,{\rm s}^{-1}\else km$\,$s$^{-1}$\fi} 
\def \halpha{H$\alpha$}
\def \ifm#1{\relax\ifmmode#1\else$\mathsurround=0pt #1$\fi} 
\def \hmpc{h\ifm{^{-1}}\,{\rm Mpc}} 
\def \etal{{\rm et al.}} 
\def \ie{{\it i.e.}} 
\def \eg{{\it e.g.}} 
\def \arcmin{\hbox{$^\prime$}} 
\def \farcs{\hbox{$.\!\!^{\prime\prime}$}} 
\def \farcm{\hbox{$.\mkern-4mu^\prime$}} 
\def \fm{\hbox{$.\!\!^{\rm m}$}} 
\def \yskip{\penalty-50\vskip3pt plus3pt minus2pt} 
\def \pp{\par\yskip\noindent\hangindent 0.4in \hangafter 1} 
\def \gtsima{$ \buildrel > \over \sim \,$} 
\def \ltsima{$ \buildrel < \over \sim \,$} 
\def \simgt{\lower.5ex\hbox{\gtsima}} 
\def \simlt{\lower.5ex\hbox{\ltsima}}
\def \bfV{{\bf V}} 
\def \bfr{{\bf r}} 
\def \bfq{{\bf q}} 
\def \bfx{{\bf x}} 
\def \bfk{{\bf k}} 
\def \bfz{{\bf z}} 
\def \bfu{{\bf U}} 
\def \nhat{\ifmmode {\hat{\bf n}}\else${\hat {\bf n}}$\fi}
\def \xhat{\ifmmode {\hat{\bf x}}\else${\hat {\bf x}}$\fi}
\def \yhat{\ifmmode {\hat{\bf y}}\else${\hat {\bf y}}$\fi}
\def \zhat{\ifmmode {\hat{\bf z}}\else${\hat {\bf z}}$\fi}
\def \kmsmpc {\kms\ {{\rm Mpc}}^{-1}}

\def \littleprime{\ifmmode{\scriptscriptstyle \prime } 
     \else{\hbox{$\scriptscriptstyle \prime$ }}\fi} 
\def \littless{\ifmmode{\scriptscriptstyle s } 
     \else{\hbox{$\scriptscriptstyle s $ }}\fi} 
\def \littlemm{\ifmmode{\scriptscriptstyle m } 
     \else{\hbox{$\scriptscriptstyle m $ }}\fi} 
\def \littlecirc{\ifmmode{\scriptscriptstyle \circ } 
     \else{\hbox{$\scriptscriptstyle \circ $ }}\fi} 
\def \littlehour{\ifmmode{\scriptscriptstyle h } 
     \else{\hbox{$\scriptscriptstyle m $}}\fi}
\def \toph{\raise .9ex \hbox{\littlehour}} 
\def \hourpoint{\hbox to 2pt{}\rlap{\hskip -.5ex \toph}.\hbox to 2pt{}} 
\def \topemm{\raise .9ex \hbox{\littlemm}} 
\def \magpoint{\hbox to 2pt{}\rlap{\hskip -.5ex \topemm}.\hbox to 2pt{}} 
\def \arcss{\raise .9ex \hbox{\littless}} 
\def \arcsspoint{\hbox to 1pt{}\rlap{\arcss}.\hbox to 2pt{}} 
\def \sdeg{\raise .9ex \hbox{\littlecirc}} 
\def \degpoint{\hbox to 1pt{}\rlap{\sdeg}.\hbox to 2pt{}} 
\def \arcmin{\raise .9ex \hbox{\littleprime}} 
\def \arcminpoint{\hbox to 1pt{}\rlap{\arcmin}.\hbox to 2pt{}} 
\def \arcsec{\raise .9ex \hbox{\littleprime\hskip-3pt\littleprime}} 
\def \arcsecpoint{\hbox to 1pt{}\rlap{\arcsec}.\hbox to 2pt{}} 
\def \plotfiddle#1#2#3#4#5#6#7{\centering \leavevmode 
\vbox to#2{\rule{0pt}{#2}} 
\includegraphics{#1}} 

\title{Shellflow. I. The Convergence of the Velocity Field at 6000 \kms}

\author{St\'ephane Courteau$^{1,6}$, Jeffrey A. Willick$^{2,6}$, 
        Michael A. Strauss$^{3,6}$, David Schlegel$^{5,3,6}$, and
        Marc Postman$^{5,6}$} 

\affil{(1) \ NRC/Herzberg Institute of Astrophysics, Victoria, BC, and \break
             University of British Columbia, Vancouver, BC}
 
\affil{(2) \ Deceased. Formerly at Stanford University, Department of Physics, Stanford, CA} 
 
\affil{(3) \ Princeton University Observatory, Princeton, NJ} 
 
\affil{(4) \ Durham University, Department of Physics, South Durham} 

\affil{(5) \ Space Telescope Science Institute, Baltimore, MD} 

\affil{(6) \ Visiting Astronomer, Kitt Peak National Observatory and
Cerro Tololo Inter-American Observatory} 
\setcounter{footnote}{5}

\bigskip
\bigskip

{\centerline{\bf Epitaph for Jeffrey Alan Willick (October 8, 1959 -
 June 18, 2000) }}

{\noindent We would like to dedicate this paper to the memory of our friend and 
collaborator Jeff Willick who died tragically a week after our paper 
was accepted for publication.  The world of astronomy lost a superb 
scientist; we also lost a very dear friend.  We will miss him deeply.}

\begin{abstract}
We present the first results from the Shellflow program, 
an all-sky Tully-Fisher (TF) peculiar velocity survey of 276 Sb$-$Sc galaxies
with redshifts between 4500 and 7000 \kms.
Shellflow was designed to minimize systematic errors 
between observing runs and between telescopes, thereby removing the
possibility of a spurious bulk flow caused by data inhomogeneity.
A fit to the data yields a bulk flow amplitude 
$V_{{\rm bulk}} = 70^{+100}_{-70}\ \kms$ ($1\sigma$ error)
with respect to the Cosmic Microwave
Background, i.e., consistent with being at rest. 
At the 95\% confidence level, the flow amplitude is $< 300\,\kms.$
Our results are insensitive to which Galactic extinction maps
we use, and to the parameterization of the TF relation.  
The larger bulk motion found in 
analyses of the Mark III peculiar velocity catalog are thus likely
to be due to non-uniformities between the subsamples making up Mark III. 
The absence of bulk flow is consistent with the study of Giovanelli
and collaborators and flow field predictions from the observed 
distribution of IRAS galaxies.
\end{abstract}


\section{Introduction}
\setcounter{footnote}{0}

It is of great cosmological importance to identify the volume of space,  
centered on the Local Group, which is at rest with respect to the 
Cosmic Microwave Background radiation (CMB).
Very large-scale density fluctuations are required to move large volumes 
of galaxies in the gravitational instability picture of structure formation.  
In standard Cold Dark Matter (CDM) cosmogonies, density fluctuations on scales
$\simgt 100\hmpc$ are very small. As a result, 
the volume of space encompassed by the nearest superclusters (Great 
Attractor, Pisces-Perseus, Coma)
is expected to be
nearly at rest with respect to the CMB, and the distribution of matter 
{\em within\/} 
this volume should explain the $\sim 600$ \kms\ motion of the 
Local Group in the CMB frame.
However, the detection of a large amplitude flow
($V_{{\rm bulk}} \simgt 700$ \kms) out to 15,000 \kms\ 
by Lauer \& Postman (1994), along with 
recent measurements of similar amplitude (although different directions)
by Willick (1999b) and Hudson \etal\ (1999),
have challenged the notion that the bulk flow on large scales is
small, and 
are pushing CDM models to the breaking point (e.g., Feldman
\& Watkins 1994; Strauss \etal\ 1995).  However, Giovanelli \etal\
(1998a,b) and Dale \etal\ (1999) find results consistent with no flow 
in their survey of field and cluster spirals out to 20,000 \kms. 

The measured bulk flow on smaller scales is also controversial. 
The most recent POTENT reconstructions (Dekel \etal\ 1999) of the
Mark III velocities (Willick \etal\ 1997; Mark III)
find a bulk velocity within 6000 \kms\ of $370 \pm 110$ \kms\ in 
the CMB frame towards Supergalactic $(L,B)=(165^\circ,\,
 -10^\circ)$\footnote{A slightly smaller bulk flow amplitude
of $305 \pm 110\kms$ is obtained if the VELMOD2 TF calibration
of Willick \& Strauss (1998) is used.}. 
Dekel \etal\ (1999) argue that this motion
is generated by the {\it external} mass distribution on very large
scales (see also Courteau \etal\ 1993).  On the other hand, Giovanelli
\etal\ (1998) find a flow consistent with zero on similar scales from
their field sample, a result consistent with the surface brightness
fluctuation data of Tonry \etal\ (2000) and SN Ia distances (Riess
2000). 

Accurate ($\simlt 150\ \kms$) 
measurement of the bulk flow at 6000 \kms\ requires
that the galaxy distance data be homogeneous and free of systematic
effects at the $2-3$\% level. This
cannot be guaranteed for datasets, such as the Mark III catalog,
that are composed of two or more independent
peculiar velocity surveys.
Indeed, Willick \& Strauss (1998) found evidence of
systematic errors in the relative zero points of the various
TF samples that make up the Mark III catalog. 
Thus, the controversy over the observed bulk flow within $60 \hmpc$ stems, 
in large part, from the difficulty of combining the various galaxy distance 
samples used in flow studies into a single homogeneous catalog.
None of the previous surveys 
extending to $60\hmpc$ sampled the {\it entire} sky 
uniformly and reduced the raw data for Northern and Southern hemisphere
galaxies using identical techniques\footnote{Earlier attempts include 
Roth (1994) and Schlegel (1995).  
The work of Giovanelli \etal\ (1998a,b) incorporates the Southern 
galaxy survey of Mathewson \etal\ (1992), but these authors claim to
have reduced the systematic offset in the calibration between the two
data sets to negligible levels.  On scales larger than 
6000 \kms, the surveys of Lauer \& Postman (1994) and Dale \etal\ (1999)
were designed in a manner analogous to Shellflow.}.

To address these issues we undertook a new
TF survey focussed on a relatively narrow redshift shell
centered at $\sim 6000\ \kms.$ 
Our survey, ``Shellflow,''
was designed to provide {\em precise\/} and {\em uniform\/} 
photometric and spectroscopic data over the whole sky,
and thus to remove the uncertainties associated with matching 
heterogeneous data sets. 
In this {\em Letter,} we report the first scientific result
from Shellflow: a reliable, high-accuracy measurement
of the bulk flow at $60\hmpc.$ In future papers 
(Willick \etal\ 2000, Paper II; Courteau \etal\ 2000, Paper III), 
we will describe the data set in greater detail and address
related scientific questions, including higher-order
moments of the flow field and
the value of $\beta\equiv \Omega_m^{0.6}/b.$

\section{Sample Selection and Observations}
 
The Shellflow sample is drawn from the Optical Redshift Survey sample
of Santiago \etal\ (1995; ORS). The ORS sample consists of all galaxies
in the UGC, ESO, and ESGC Catalogs  with
$m_{\rm{B}} \leq 14.5$ and $|b| \geq 20^\circ.$ 
We selected all non-interacting Sb and Sc galaxies
in the ORS with redshifts between  
4500 and 7000\footnote{We actually define 
three subsamples complete in that range with different definitions 
of redshift: measured in the Local Group frame, the CMB frame, and 
after correction for peculiar velocities according to 
the {\sl IRAS\/} model of Yahil \etal\ (1991) with $\beta = 1$; if we chose 
only one of them, the sample would decrease in size by 20\%.}
\kms,  inclinations between  
$45^\circ$ and $78^\circ$, and with Burstein-Heiles (1982; BH) 
extinctions $A_B\leq$ 0\magpoint30.  
All galaxies were inspected on the Digitized POSS scans 
(Lasker 1995) to determine their morphological types and inclinations;
those galaxies with bright
foreground stars and tidal disturbances were excluded, yielding 
a final sample of 297 Shellflow galaxies.
No pruning was done of galaxies not matching idealized 
morphologies beyond the restriction on Hubble type and inclination. 

We collected V and I-band CCD photometry\footnote{This paper 
focuses solely on I-band imaging.
The V-band data will be used in future papers to verify extinction corrections
and photometric errors.} and \halpha\ rotation curves 
between March 1996 and March 1998 using NOAO facilities; this paper reports results
based on the 276 galaxies for which we obtained high-quality imaging
and spectroscopic data. 
Data taking and reduction techniques follow the basic guidelines of 
previous optical TF surveys (\eg\ Schlegel 1995; Courteau 1996, 1997).
The V and I-band images were obtained at the CTIO and KPNO 0.9m 
telescopes.  The photometric calibration is based on the Kron-Cousins 
system; data taken on nights with standard star photometric scatter 
greater than 0\magpoint02 were excluded.  The Kron-Cousins 
system also allows direct matching with the two largest I-band TF 
samples to date (Mathewson \etal\ 1992, Giovanelli \etal\ 1998). 
The \halpha\ spectroscopy was obtained mostly in photometric 
conditions 
with the RC spectrographs at the CTIO and KPNO 4m telescopes.
Typical integrations were $\sim 900$s and $\sim 1800$s for imaging 
and spectroscopy respectively.  The position angle of each galaxy,
for orientation of the spectrograph slit, was inferred from surface 
photometry off the Digitized POSS scans.
As some of the spectroscopy was obtained before the CCD imaging runs,
we were unable to use the CCD data to determine the orientation.
However, a posteriori checks has shown no systematic offset, and 
tiny scatter, between the position angles measured from the DPOSS
and CCD images (Courteau et al. 2000).

Forty-one galaxies were imaged at both CTIO and KPNO, and we have
repeat imaging from a given telescope for a third of our sample. 
In addition, we observed 27 galaxies spectroscopically from both CTIO 
and KPNO, and obtained duplicate spectra from a given telescope for 38 
galaxies.  The total magnitudes and rotational line widths reproduce to
within 0\magpoint06 and 3 \kms\ (rms deviations) respectively, with no
systematic effects seen between hemispheres or between runs. 
All data reduction was done independently by Courteau and
Willick using different software and methodology; the results between
the two agree to within the errors quoted above. 
The small random and systematic errors of the Shellflow
data meet our requirements for a bulk flow measurement
with overall rms error $\simlt 150\ \kms.$

Systematic differences in photometric scale length exist between 
Courteau's and Willick's reduced data sets
due to different methods of luminosity profile fitting. 
These differences affect the TF relation because our extrapolated 
magnitudes depend on the inferred disk profile and we measure
the rotation velocity at a fixed multiple of scale lengths (\S~3).
Use of Willick's ``moment method'' (Willick 1999a) for 
determining scale lengths leads to a small but 
significant surface brightness dependence of the TF relation, whereas
use of Courteau's fitted exponential scale lengths (Courteau 1996, 
Courteau \& Rix 1999) does not. 
We will discuss these issues in detail in Paper III.  However, 
the bulk motions we find are virtually identical whether we adopt
Courteau's or Willick's reduced data set for the analysis; for the
remainder of this paper we use Willick's reduced data set, for
which a somewhat smaller TF scatter is obtained.

\section{Analysis and Results} 

Following Lauer \& Postman (1994) and Willick (1999b), we
calibrate the distance indicator relation with the sample 
itself, fitting for the velocity field simultaneously; this obviates 
the need to tie the sample to external TF calibrators such as clusters. 

We adopt the ``inverse'' form of the TF relation (minimizing
velocity-width rather than magnitude residuals)
for which Malmquist and selection bias effects 
are negligible (Schechter 1980, Strauss \& Willick 1995). 
We write the I band inverse TF relation 
\be
\eta = -e(M_I-D)  - \gamma (\mu_I - 18.6) + \beta(c-2.4)\,.
\label{eq:TFrel} 
\ee
Here $M_I$ is absolute magnitude, $\mu_I$ effective surface brightness,
$c$ a logarithmic concentration index (as defined in Willick 1999a),
$D$ and $e$ are the zero point and slope, respectively, of the TF relation,
$\gamma$ and $\beta$ represent the possible additional dependences
on surface brightness and concentration; and $\eta\equiv\log(2v_{{\rm rot}})
-2.5$ is the velocity width parameter. As noted in \S~2, the
dependences represented by $\gamma$ and $\beta$ are small, and we obtain
virtually identical flow results if we set $\gamma = \beta \equiv 0.$ 
We obtain $v_{{\rm rot}}$ as follows: first, we fit the
\halpha\ rotation curve (RC) to a parameterized functional form
(we use a modified arctangent fit, but the exact parameterization
is not important), yielding a smooth RC $v(R).$ We then evaluate
the RC at a galactocentric radius $f_s R_e$---i.e., we
take $v_{{\rm rot}}=v(f_s R_e)$---where $R_e$ is the
exponential scale length measured from the photometric
profile. We treat the quantity $f_s$ as a free parameter in
the fit; we ultimately find $f_s \approx 1.7,$ in rough agreement
with earlier work by Courteau (1997) and
Willick (1999a), although the precise value of $f_s$ depends
on the particular scale length used, as we discuss in
detail in Paper III. 

The absolute magnitude is obtained from the usual expression
$M_I = m_I + 5\log d({\rm Mpc})+25,$ where $m_I$ is the
measured I band apparent magnitude corrected
for Galactic and internal extinction (see below), 
and the distance is given
by a Hubble expansion plus bulk flow model,
\be
d ({\rm Mpc}) = \frac{1}{H_0} \left(cz - \bfV_B\cdot\nhat\right) \,, 
\ee
where $cz$ is the redshift, either in the CMB or Local Group frame, $\nhat$ is a unit
vector in the direction of the galaxy, and $\bfV_B$
is a bulk flow vector. We adopt 
$H_0=100\,\kmsmpc,$ but emphasize that the value
of the Hubble constant affects only the
zero point $D$ of the TF relation, not
the bulk flow result. 
The three Cartesian components of $\bfV_B$
are additional free parameters in the maximum-likelihood procedure.
The Galactic extinctions used to correct the apparent
magnitudes are obtained from the maps of Schlegel, Finkbeiner,
\& Davis (1998). However, our flow results
are virtually unchanged if we use the BH extinctions, 
as we discuss in detail in Paper III. For the internal
extinctions we adopt the simple formula 
$A_I^{int}= - 0.5 \log(1-\varepsilon),$
where $\varepsilon$ is the ellipticity of the galaxy image
as determined from surface photometry. 
We will justify this extinction 
formula in Paper III by showing that it leads
to TF residuals that are uncorrelated with inclination.

The maximum likelihood solution is obtained by minimizing 
${\cal L} \equiv -2\sum_i \ln P_i,$ where the sum runs over all
sample objects, and the probability for a single object is
given by
\be
P_i = \frac{1}{\sqrt{2\pi}\sigma_{\eta,i}}\,\exp\left[-\frac
{(\eta_{i,obs} - \eta_{i,pred})^2}{2\sigma_{\eta,i}^2}\right] \,,
\ee
where $\eta_{i,obs}$ and $\eta_{i,pred}$ are the observed
and predicted (from Eq.~\ref{eq:TFrel}) velocity width
parameters. We model the inverse TF scatter as
$\sigma_\eta^2 = \sigma_{\eta,{\rm int}}^2 +
\sigma_{\eta,{\rm phot}}^2 + \sigma_{\eta,v}^2,$
i.e., as a quadrature sum of intrinsic scatter,
the effect of photometric measurement errors on $\eta$
(i.e. apparent magnitude measurement errors times
the inverse TF slope, which affect $\eta_{pred},$
plus the effects of scale length
and inclination measurement errors, which affect
$\eta_{obs}$), and raw velocity width
measurement errors. We obtain $\sigma_{\eta,{\rm phot}}$
and $\sigma_{\eta,v}$ from repeat observations,
which enables us to treat the instrinsic scatter $\sigma_{\eta,{\rm int}}$
as a free parameter in the fit. In Paper II, we discuss these issues
in greater detail; for now we note merely that there is no covariance
between TF scatter and bulk flow.

There are therefore nine free parameters in
the maximum likelihood fit: six TF parameters ($D,$
$e,$ $\sigma_{\eta,{\rm int}},$ $\gamma,$ $\beta,$
and $f_s$), plus the three
components of the bulk flow vector.
In Table 1 we present the best-fitting
values of the six TF parameters.
The zero point and slope are comparable to those of recent 
I-band studies (\eg\ Giovanelli \etal\ 1997).  
The intrinsic scatter divided by the inverse TF slope yields
an equivalent {\em forward\/} intrinsic TF scatter of
$0.25$ mag, similar to earlier estimates of the intrinsic
TF scatter (e.g., Willick \etal\ 1996;  Giovanelli
\etal\ 1997; Willick 1999b). The values of $\gamma$ and $\beta,$
which describe the surface-brightness and concentration-index
dependences of the TF relation are, as noted above, small,
while $f_s\simeq 1.7$ is similar to, though somewhat smaller
than, the corresponding values obtained by Courteau (1997)
and Willick (1999a,b). 
Contrary to Giovanelli \etal\ (1995), whose TF relation is also based
on I-band imaging, we find no evidence for a luminosity dependence
of internal extinction for this Shellflow sample.  This result is 
consistent with our analysis of the Mathewson I-band sample in
Willick \etal\ (1996).  Further details about TF dependences will 
be addressed in Paper III.

\begin{table}[!ht]
\begin{center}
\centerline{{\sc Table 1: I-band TF Fit Parameters}}
\vskip 4pt
\begin{tabular}{cccccc} \hline\hline
$D$ & $e$ & $\sigma_{\eta,{\rm int}}$ & $\gamma$ & $\beta$ & $f_s$ \\ \hline
$-20.96$ &$0.124$ & $0.031$ & $0.044$ & $0.034$ & $1.69$ \\ \hline
\end{tabular}
\end{center}
\end{table}

\begin{figure}[ht!]
\includegraphics[scale=0.60]{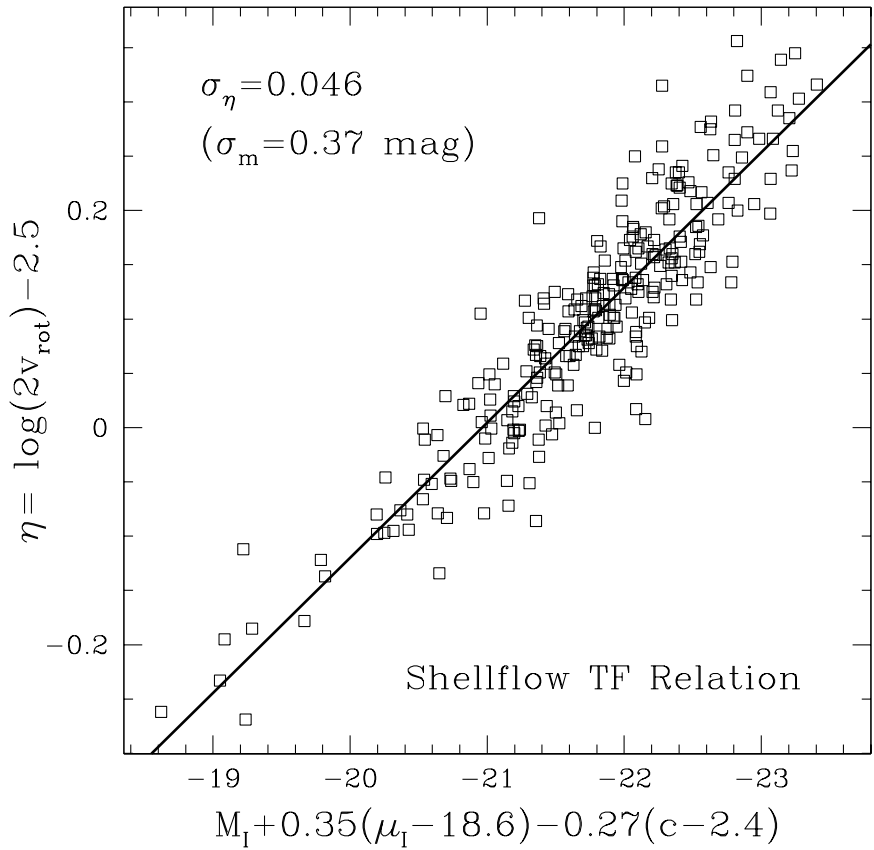}
\caption{{\small
 I-band Tully-Fisher relation for the Shellflow sample.
 The absolute magnitudes $M_I$ are corrected for the
 small surface-brightness and concentration-index
 dependences of the TF relation (see Eq.~1).
 The solid line is the best-fit TF relation whose
 parameters are given in Table~1. Also indicated are
 the overall inverse TF scatter $\sigma_\eta,$ and the
 corresponding forward TF scatter $\sigma_m=\sigma_\eta/e,$
 where $e$ is the TF slope.}}
\label{fig:ShellTF}
\end{figure}

In order to exhibit the multiparameter TF relation graphically,
we define effective absolute magnitudes by
$M_I^{{\rm eff}} = M_I + (\gamma/e)(\mu_I-18.6) - (\beta/e)(c-2.4).$
From Eq.~(\ref{eq:TFrel}), $\eta$ depends linearly on $M_I^{{\rm eff}}$
with slope $e.$ In Figure~\ref{fig:ShellTF} we plot
$\eta$ versus $M_I^{{\rm eff}}$ for the Shellflow sample;
the absolute magnitudes themselves are obtained using the
best-fit bulk flow model (see below).
Multiply-observed galaxies are shown as single points at their
average spectroscopic and photometric parameters. The straight line
shows the best fit TF relation from Table~1. Also
indicated on the Figure are the values of the overall inverse TF scatter
$\sigma_\eta$ and the corresponding overall forward scatter $\sigma_m
=\sigma_\eta/e.$ By subtracting the intrinsic TF scatter in quadrature from
these values, one finds that the contribution of 
measurement errors to the overall
scatter is comparable to the intrinsic scatter, $\sim 0.25$--$0.30$ mag.

\begin{figure}[ht!]
\includegraphics[scale=0.60]{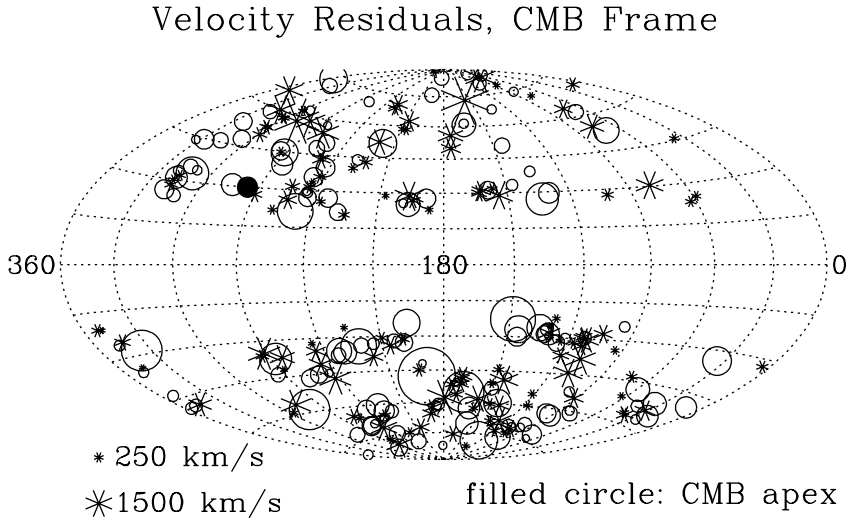}
\caption{{\small
 Apparent peculiar velocity residuals (TF residuals converted
 into velocities) of the
 Shellflow galaxies for a pure Hubble flow fit in the CMB frame.  
 Point size is proportional to the velocity amplitude,
 with fiducial values indicated at the lower left.
 The circles and asterisks represent 
 inflowing and outflowing objects, respectively.  
 The filled circle shows the
 location of the CMB dipole (Kogut \etal\ 1993).}} 
\label{fig:CMBresid}
\end{figure} 

The best-fitting values of the three components
of the bulk flow vector are listed in Table 2.  
Velocities are in units of $\kms,$ and the Cartesian components
are taken with respect to the Galactic coordinate system:
$\xhat=\cos b\,\cos \ell,$ $\yhat=\cos b\,\sin \ell,$
$\zhat=\sin b.$
We list results for both CMB and Local Group (LG)
frame fits.  The
$1\,\sigma$ errors on each velocity component are also indicated.
These  errors are derived by running through a range of values, 
$\pm 250\,\kms$ relative to the best-fit value, of each component,
and holding it fixed while maximizing likelihood relative
to the other two velocity components; the $1\sigma$ errors
are the values of the velocity components 
for which the likelihood statistic
$\cal L$ differs by one unit from its minimum value.\footnote{We
do not vary the TF parameters as well in this exercise, as there
is essentially no covariance between them and the velocities;
a pure Hubble flow fit yields essentially the same TF parameter values.} 
We similarly vary all three velocity components simultaneously to find
the the errors on the flow amplitude $V_{{\rm bulk}}=
\sqrt{V_x^2+V_y^2+V_z^2}$.
In this
way we find that $V_{{\rm bulk}}=0$ is within $1\sigma$ of the
best fit, while $V_{{\rm bulk}}\ge 170\,\kms$ corresponds to
$\Delta{\cal L} \ge 1.$  We thus obtain our $1\sigma$ bounds
on the bulk flow amplitude as $V_{{\rm bulk}}=70^{+100}_{-70}\,\kms$.
We similarly find that $V_{{\rm bulk}} \ge 300\,\kms$
corresponds to $\Delta{\cal L} \ge 4,$ indicating that
flow amplitudes $\ge 300\,\kms$ are
ruled out at the $2\sigma$ (95\% confidence) level.
The monopole of the velocity field couples with the dipole moment of
the sample distribution because we are using the sample itself to
calibrate the TF relation (Lauer \& Postman 1994).  We estimate the
amplitude of this geometric bias on the dipole to be of the order of
50 \kms.  A detailed error analysis based on Monte Carlo simulations,
in which covariance among the velocity components is
fully explored, will be presented in Paper II.  

\begin{table}[!ht]
\begin{center}
\centerline{{\sc Table 2: Bulk Flow Solutions }}
 \vskip 4pt
 \begin{tabular}{rrr|l}\hline\hline  
 V$_z$ \  & V$_x$ \  & V$_y$ \  & Frame \\
 \noalign{\vskip 3pt}
 \hline  
 \noalign{\vskip 5pt}
 $  56 \pm 75$&$ -38 \pm 115$&$  28 \pm 115$&CMB   \\  
 $-256 \pm 75$&$ -96 \pm 115$&$ 569 \pm 110$&LG    \\ \hline  
 \end{tabular}  
\end{center}
\end{table}

In Figure~\ref{fig:CMBresid} apparent peculiar velocities
in the CMB frame are plotted in Galactic coordinates.
These velocities are calculated as $v_p=cz \frac{\ln 10}{5e}\delta\eta,$
where $\delta\eta$ is the TF residual in a pure Hubble flow model.
The symbol types and sizes indicate the sign and amplitude
of the velocities. Inflowing and outflowing objects
are well-mixed at all positions on the sky, indicating the
absence of a coherent flow, as our likelihood fits confirm.
Most velocity amplitudes are $\simlt 2000\,\kms,$ corresponding
approximately to a $2\sigma$ TF residual at 6000 \kms, and thus are not
individually significant.

If the Shellflow sample is 
at rest in the CMB, we expect to see the reflex of the
LG motion through the CMB when we analyze the flow
in the LG frame. This is indeed what we see,
as the second row of Table~2 shows. The flow amplitude is
$631\,\kms,$ and the flow vector is directed
towards $\ell=89\degs,$ $b=-27\degs.$ This amplitude is
very nearly the same as, and the direction is almost precisely
opposite from, the vector of the
LG motion as
determined from the CMB dipole anisotropy (e.g., Kogut \etal\ 1993).

\section{Discussion}

The results we have presented here are in broad agreement with
other recently reported results on the flow field in the
local universe.  These include the analyses of the SCI and SFI
TF samples (Giovanelli \etal\ 1998a,b), who find V$_{\rm bulk}=200\pm65$
\kms\ within 6500 \kms\ and no motion for shells farther than 5000 \kms;
a similar analysis by Dale et al. (1999) who find no significant
motion of clusters between 5000 and 20000 \kms; as well as work
from Tonry \etal\ (2000), who obtain V$_{\rm bulk}=289\pm137$ \kms\ 
at 3000 \kms\ from surface brightness fluctuation data, and Riess (2000),
who finds no measurable bulk flow in the CMB frame from a sample
of 44 SNe Ia with an average depth of 6000 \kms.  
Taken together these results suggest that by a distance of $60\hmpc$,
we are seeing a convergence of the flow field to the CMB frame, as is
predicted by the observed distribution of IRAS galaxies (Strauss \etal\
1992; Schmoldt \etal\ 1999; Rowan-Robinson \etal\ 2000).
While the data for more distant samples remain ambiguous,
with several claims of large amplitude flows on scales
$\simgt 100\hmpc,$ the results within $60\hmpc$ cast serious doubt
on these claims. If, as abundant evidence suggests, the universe
monotonically approaches homogeneity on ever larger scales,
it is difficult to see how $\simgt 600\,\kms$ bulk flows on
$\simgt 100\hmpc$ scales can be reconciled with
negligible bulk flow on a scale half as large. From this
perspective it seems likely that the results
of Lauer \& Postman (1994), Willick (1999b), and
Hudson \etal\ (1999) are due, at least in part,
to subtle and small systematic effects. 

In summary, we find no significant motion of a shell of galaxies
centered at 6000 \kms, as seen in the CMB frame. Equivalently, from 
the vantage point of the LG frame, we see a motion equal in amplitude
and opposite in direction to the motion of the LG through the CMB.
Our results are insensitive to whether we adopt the BH or the SFD
reddenings, as well as to the parameterization of the TF relation.
Future papers will present the spectroscopic and photometric data, 
give a detailed account of our TF analysis, including tests for a 
surface-brightness dependence of the TF relation, consider 
higher-order moments of
the velocity field, and compare with the IRAS-predicted velocity
field, following the methods of Davis, Nusser, \& Willick (1996)
and Willick \& Strauss (1998), to estimate
$\beta=\Omega_m^{0.6}/b.$  We will also use the Shellflow sample
to recalibrate and homogeneously merge the major TF catalogs
out to 6000 \kms, including Mark III and SFI (Haynes \etal\ 1999).
Such a future superset of existing TF catalogs, based on a 
reliable, all-sky calibration, will provide a powerful tool for
studying the velocity and density fields in the local universe.

\acknowledgments We wish to thank various students and postdocs who
have contributed to the Shellflow reductions: Shelly Pinder and
Yong-John Sohn in Victoria, and Josh Simon, Felicia Tam, and Marcos
Lopez-Caniego at Stanford.  SC acknowledges support from the National
Research Council of Canada, JAW from Research Corporation and NSF grant
AST96-17188, and MAS from Research Corporation and NSF grant AST96-16901.

\pagebreak

\end{document}